\documentclass{aa} 

\usepackage{natbib}
\bibpunct{(}{)}{;}{a}{}{,} 

\usepackage{array}
\usepackage{makecell}
\usepackage{placeins}
\usepackage{graphicx}
\usepackage{deluxetable} 
\usepackage{balance}
\usepackage[colorlinks=true,linkcolor=blue,citecolor=blue,urlcolor=blue]{hyperref}
\usepackage{txfonts}

\begin{document} 

   \title{Flaring together: A preferred angular separation between sympathetic flares on the Sun}
   
   \author{L-S. Guité
          \inst{1},
          A. Strugarek\inst{2}
          \and
          P. Charbonneau\inst{1}
          }

    \institute{Physics Department, Université de Montréal, CP 6128 Centre-Ville, Montréal, Qc H3C-3J7, Canada \\
    \email{louis-simon.guite@umontreal.ca}  
    \and
    Université Paris-Saclay, Université Paris Cité, CEA, CNRS, AIM, 91191, Gif-sur-Yvette, France\\
    }

\abstract{Sympathetic solar flares are eruptions that occur nearby in space and time, driven by an apparent interaction between the active regions in which they are triggered. Their statistical existence on the Sun has yet to be firmly established.}
{The main goal of this paper is to identify a statistical signature of sympathetic flares, characterize their properties and determine a potential mechanism driving their interaction.}
{We perform a statistical analysis of a large number of flares observed by the Atmospheric Imaging Assembly (AIA) onboard the Solar Dynamics Observatory (SDO), the Reuven Ramaty High Energy Solar Spectroscopic Imager (\textit{RHESSI}) and the Spectrometer Telescope for Imaging X-rays (STIX) on Solar Orbiter during solar cycle 24 and 25. We examine the spatiotemporal distribution of consecutive flare pairs across solar cycle phases and hemispheres, along with the propagation velocity of potential causal interactions and the relationship between flare magnitudes.}
{We observe an excess of hemispheric flares separated by about 30 degrees of longitude and triggered in less than 1.5 hours from each other. This peak in angular separation varies with the solar cycle phase and hemisphere. Moreover, we identify a deficit of transequatorial events separated by 25-30 degrees in latitude and less than 5 degrees in longitude, a phenomenon we term \textit{unsympathetic flares}.}
{We provide strong statistical evidence for the existence of sympathetic flares on the Sun, demonstrating that their occurrence rate reaches approximately 5\% across the three instruments used in this study.  Additionally, we propose an interpretation of the observed angular scale of the sympathetic phenomenon, based on the separation between magnetic field line footpoints derived from potential field source surface extrapolations.} 


\titlerunning{Flaring together}
\authorrunning{Guité, L-S., et al.}

\maketitle

%

\section{Introduction}
\par Solar eruptions are said to be \textit{sympathetic} when they are triggered almost synchronously but originate from distinct regions on the solar surface, presumably caused by a physical interaction between them. Since the first investigation of this phenomenon by \citet{Richardson1951}, sympathetic eruptions have been studied either statistically or case-by-case. From a statistical standpoint, several studies have found an overabundance of nearby flares at short waiting times compared to a Poisson distribution, which would be expected for sympathetic flares \citep[e.g.][]{Fritzova-Svestkova1976, Pearce1990, Moon2002, Wheatland2006, Schrijver2015}. However, the unambiguous statistical signature of these synchronous eruptions has been debated because other studies found no evidence for such sympathy
\citep[e.g.][]{Biesecker2000, Moon2001, Moon2003}.
\\\\
\begin{figure*}[ht]
    \centering \includegraphics[width=2\columnwidth]{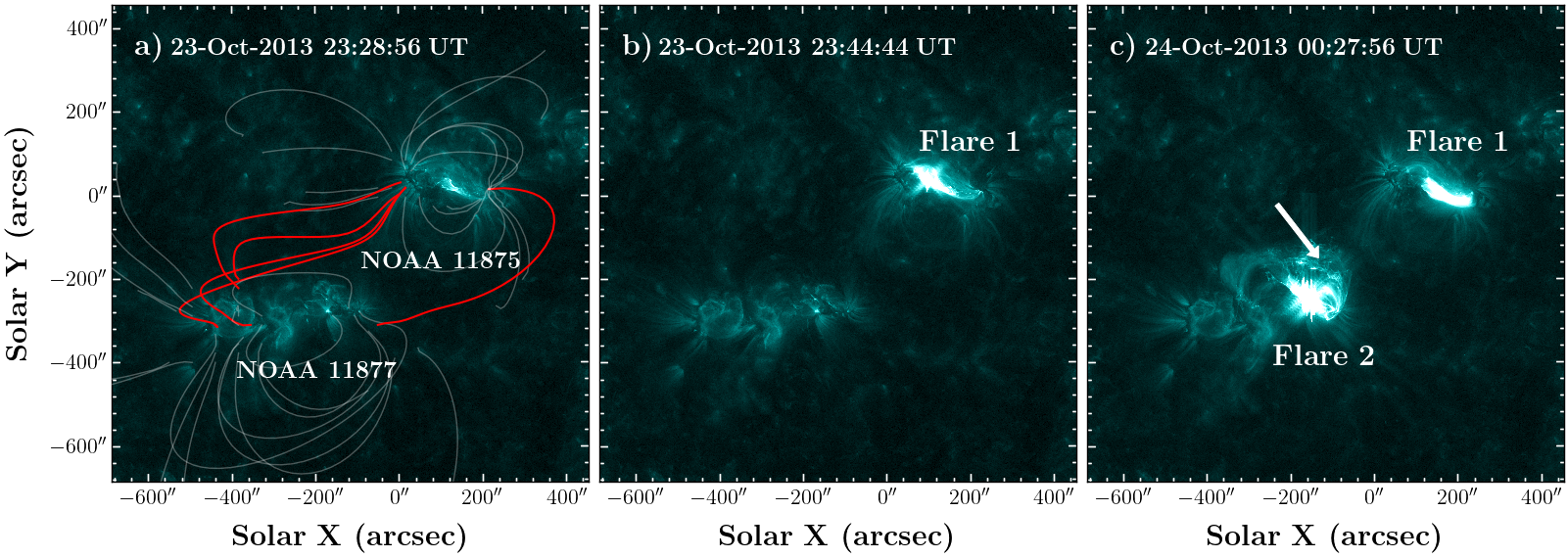}
\caption{Example of a candidate sympathetic event observed by SDO/AIA at 131 Å on October 23-24, 2013, with panels (a) to (c) showing the chronological evolution of the flaring regions. In panel (a), the white and red lines represent a PFSS extrapolation made using the \texttt{pfsspy} code from \citet{Stansby2020}, with the source surface radius at $r_{SS} = 2.5$R$_{\odot}$. In panel (c), the white arrow indicates the location of a loop brightening connecting the two active regions (NOAA 11875 and NOAA 11877). An animated version of this figure is available online.}
\label{fig:example}
\end{figure*}
\noindent Apparent causal links have been observed in a variety of eruptive events, such as flares \citep[e.g.][]{Bumba1993, Bagala2000, Wang2001, Liu2009}, coronal mass ejections (CMEs) \citep[e.g.][]{Moon2003, Cheng2005, Jiang2008, Jiang2009}, erupting filaments \citep[e.g.][]{Jiang2011, Shen2012, Yang2012, Joshi2016, Wang2016, Li2017, Wang2018}, or even coronal bright points \citep[e.g.][]{Pucci2012, Zhang2013}, based on factors like the location and timing of eruptions, the observed emission or the magnetic configuration of the system. Based on observations in different wavelengths, several mechanisms have been suggested to explain sympathetic eruptions, such as common magnetic field evolution from photospheric convective motions \citep{Bumba1993}, ejection of plasma loops visible through X-ray emission between regions \citep{Gopalswamy1999},
heat conduction along coronal loops \citep{Changxi2000} or mass transfer from a sweeping closed-loop surge \citep{Wang2001}. Another promising explanation involves the direct magnetic interaction, in the form of magnetic reconnection of field lines, between neighboring regions or within a single region \citep[e.g.][]{Bagala2000, Cheng2005, Jiang2008, Liu2009, Jiang2009}. For example, based on the potential field source surface (PFSS) extrapolation of magnetic field lines surrounding two filaments, \cite{Shen2012} argued that the first filament eruption was caused by the removal of confining field above it from external magnetic reconnections. The redistribution of newly reconnected field line then destabilized the second filament which later erupted, suggesting sympathy. Such weakening of stabilizing magnetic fields overlying a filament by another erupting filament has also been observed by several authors (e.g. \citealt{Jiang2011}; \citealt{Yang2012}; \citealt{Wang2016}; \citealt{Li2017}; \citealt{Wang2018}; \citealt{Song2020}; \citealt{Hou2020}).
Moreover, \citet{Schrijver2011} found that during the 2010 August 1st twin filament eruption event, all regions where a flare or eruption originated from were connected by a system of separatrices, separators, and quasi-separatrix layers. Similar conclusions were reached by \citet{Titov2012} and \citet{Titov2017}. This reinforces the idea that the state of the corona's global magnetic field and confining fields plays a key role for the trigger of sympathetic eruptions.
\\\\
These findings are supported by magnetohydrodynamic (MHD) simulations of various magnetic configurations. For instance, \citet{Ding2006} and \citet{Peng2007} modeled an octupolar and quadrupolar magnetic configuration respectively to show that the stability of magnetic flux ropes or arcades against eruption was closely related with the activity of nearby magnetic systems. Additionally, both \citet{Torok2011} and \citet{Lynch2013}  investigated the occurrence of sympathetic eruptions in the context of a pseudo-streamer (PS) configuration. Even though they used different mechanisms to disrupt the flux ropes within the PS, i.e. an outside flux rope eruption \citep{Torok2011} and ideal footpoint shearing along the PS arcade polarity inversion line \citep{Lynch2013}, they both found that such a magnetic configuration was favorable to sympathetic eruptions and subsequent CMEs. The long-range impacts of these CMEs on nearby active regions were explored by \citet{Jin2016}, who indeed found that the propagating CME flux rope could reconnect with surrounding regions, thus reducing their magnetic confinement and making them more susceptible to eruptions.
\\\\
In Figure \ref{fig:example}, we illustrate a candidate sympathetic event observed on October 23-24, 2013 by the Atmospheric Imaging Assembly (AIA; \citealt{Lemen2012}
) on board the Solar Dynamics Observatory (SDO; \citealt{Pesnel2012}) with the 131$\AA$ wavelength filter. The flares were triggered in active regions NOAA 11875 and NOAA 11877. Panels (a) to (c) show the temporal evolution of the flaring regions. In panel (a), the white and red lines represent a PFSS made using the \texttt{pfsspy} code from \citet{Stansby2020}, with the source surface radius at $r_{SS} = 2.5$R$_{\odot}$. The boundary conditions at the photosphere were set using a magnetogram from the Global Oscillation Network Group (GONG; \citealt{Harvey1996}) for 2013-10-23T23:14:00. The field lines closely align with the direction inferred from the plasma and coronal structures in the AIA images at 131$\AA$. The red lines clearly indicate magnetic connectivity between the two active regions. In panel (c), the white arrow indicates a loop brightening between the two active regions following the flare from NOAA 11877 at 2013-10-24T00:27:56. Given the angular separation of approximately $23^{\circ}$ between the two flares and a time interval of about 50 minutes between the onset of the first and second flare, this corresponds to a propagation velocity of $v \approx 100$ km/s for the mechanism that would be responsible for the sympathetic flaring.
\\\\
The aim of this paper is to establish the characteristics of a potential causal link between sympathetic flares by conducting a statistical analysis of a large number of flares using three different instruments. More precisely, we aim to provide statistical properties of sympathetic events such as the hemispherical and transequatorial occurrences, relative frequency compared to classical flares and distribution in separation angles. The paper is structured as follows: Section \ref{Identification} outlines the methodology for identifying sympathetic flaring events and presents evidence of their statistical signature. In Section \ref{properties}, we examine the properties and characteristics of the candidate sympathetic flares, while Section \ref{sec:instruments} compares results across various instruments. A list of identified candidate sympathetic flares is provided in Section \ref{sec:list_candidate}. Section \ref{sec:interpretation} discusses potential physical mechanisms that may explain the observed sympathetic flare signal and summarizes our findings. Finally, Appendix \ref{sec:catalog} details the three flare lists used in this study, and Appendix \ref{sec:AR} presents an analysis of the spatial distribution of active regions to support the detection of sympathetic flares.


\section{Sympathetic flare identification} \label{Identification}

\subsection{Flares catalogs}
We use the flare catalogs from three instruments to have a temporal coverage over multiple solar cycles and to maximize the robustness of a potential statistical detection of sympathetic flares. Specifically, we use flare lists from SDO/AIA, the Reuven Ramaty High Energy Solar Spectroscopic Imager (\textit{RHESSI}; \citealt{lin2002}), and the Spectrometer Telescope for Imaging X-rays (STIX; \citealt{Krucker2020}) on board Solar Orbiter \citep[][]{Muller2020}. The numbers of flares from the initial lists and those used in our study for each instrument are shown in Table \ref{table:nflares}. The rationale for using a subset of the initial flare lists, along with detailed information about each list, is provided in Appendix \ref{sec:catalog}. 
\begin{table}[ht]
    \centering
    \scriptsize
    \caption{Number of Flares Detected by Instrument \label{table:nflares}}
    \begin{tabular}{lccc}
        \hline
        \hline
        \vspace{-0.2cm} \\ \footnotesize Instrument & \hspace{-0.5cm} \makecell{Number of flares \\ (Original list)} & \makecell{Number of flares \\ (Used in this work)} & \makecell{Number of \\ sympathetic flares} \vspace{0.05cm}\\
        \hline
        \vspace{-0.2cm} \\ \footnotesize SDO/AIA$\,^{1}$ & \footnotesize 16,270 & \footnotesize 13,263 & \footnotesize 922 (7\%) \\ 
        \vspace{-0.2cm} \\ \footnotesize \textit{RHESSI}$\,^{2}$ & \footnotesize 104,305 & \footnotesize 20,172 & \footnotesize 705 (3\%)\\ 
        \vspace{-0.2cm} \\ \footnotesize Solar Orbiter/STIX$\,^{3}$ & \footnotesize 17,031 & \footnotesize 8,306 & \footnotesize 322 (4\%) \\
        \hline
    \end{tabular}
    \tablebib{
(1) \citet{vanderSande2022} (2) 
\url{https://hesperia.gsfc.nasa.gov/hessidata/dbase/hessi_flare_list.txt}
    \, \,\,\,(3) \url{https://github.com/hayesla/stix_flarelist_science}}
\end{table}

\begin{figure}[h]
    \centering 
    \includegraphics[width = \columnwidth]{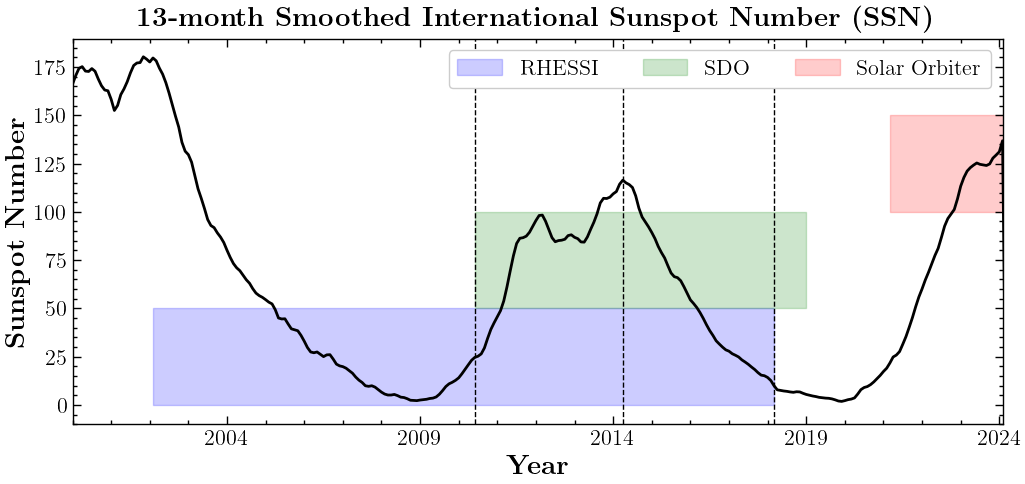}
    \caption{Temporal coverage of each flare list considered in this study, i.e. \textit{RHESSI} (blue), SDO/AIA (green) and Solar Orbiter/STIX (red). Plotted on top is the 13-month smoothed sunspot number showing the activity cycle of the Sun, downloaded from \cite{sidc}, Royal Observatory of Belgium. The vertical black dashed lines mark the time intervals referenced in Section \ref{SolarCycle}, \ref{sec:Transequatorial}, \ref{sec:sdo_rhessi} and \ref{sec:sdo_rhessi_stix} used to divide the solar cycle into its rising and decaying phases, specifically at 2010-06-01, 2014-04-01, and 2018-03-01.}
    \label{fig:SSN_TimeWindow}
\end{figure}

\noindent Note that we do not use the GOES flare list because it lacks spatial information for a sufficient number of flares, which is essential for this study. For simplicity, the main analysis is made using SDO/AIA flares except when stated otherwise. The results of the different datasets are then compared in Section \ref{sec:instruments}. The three instruments used in this study observed at different times over the past 22 years, covering solar cycles 23, 24, and the current cycle 25. To better visualize the temporal overlap of these instruments and the corresponding phases of the solar cycle during their observations, we present the time intervals for each flare list alongside the 13-month smoothed sunspot number as a proxy for solar activity in Figure \ref{fig:SSN_TimeWindow}.
\\
We initially included flares observed by Hinode and its three instruments (SOT; \citealt{Tsuneta2008}, XRT; \citealt{Golub2007} and EIS; \citealt{Culhane2007}), but decided to remove them from our analysis for the following reasons. First of all, this list had significantly fewer events at short waiting times compared to other instruments, which was essential for our study focusing on flares separated by small time intervals. Secondly, the different fields of view and observation windows of the instruments resulted in missed events, including some strong flares, when correlating with other flare lists. For example, some candidate sympathetic flares observed by SDO/AIA were missing from the Hinode flare list. Consequently, we included data only from SDO/AIA, \textit{RHESSI}, and Solar Orbiter/STIX.

\subsection{Waiting time and angular separation}
\begin{figure}[h!]
    \centering \includegraphics[width=0.7\columnwidth]{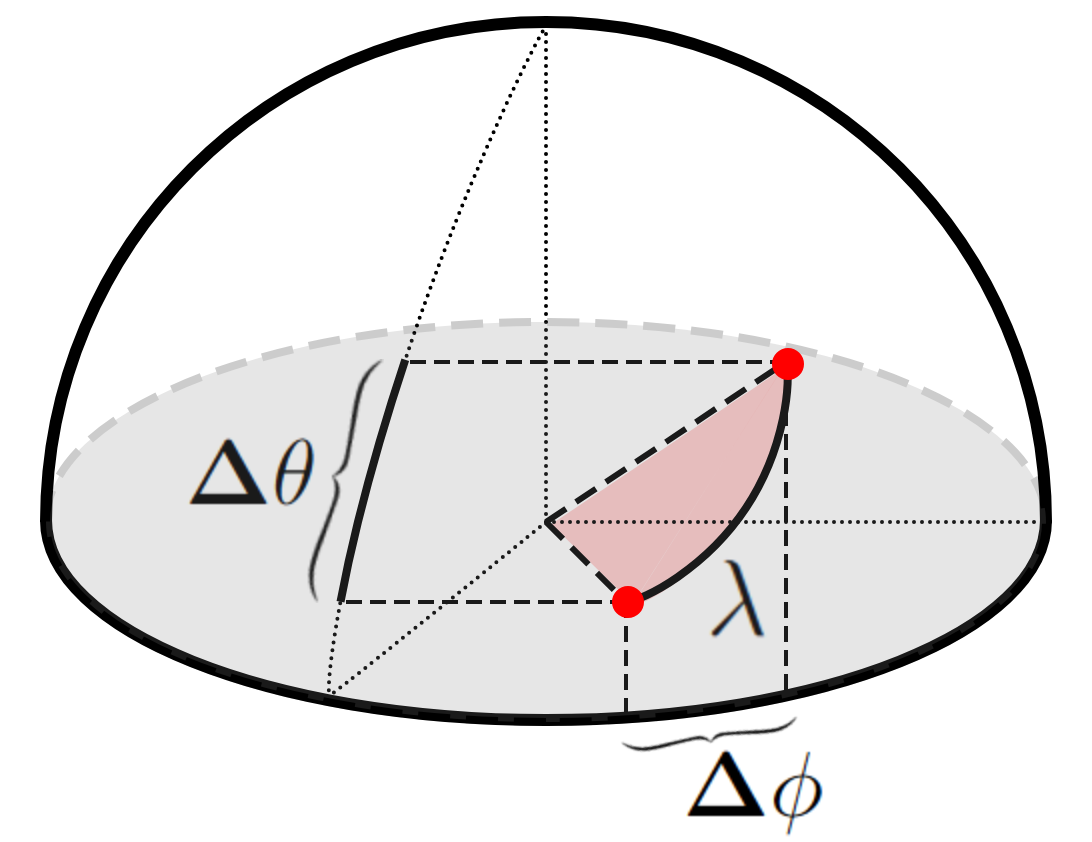}
    \caption{Schematic of the angular separation in latitude ($\Delta\theta$), longitude ($\Delta\phi$) and total angle ($\lambda$) between two flare locations marked as the two red dots.}
    \label{fig:sphere}
\end{figure}

The statistics of solar flares are often described by a non-stationary Poisson distribution, where each flare is considered independent from the previous one \citep[e.g.][]{Wheatland2000,Aschwanden2010}. However, considering that separate active regions can influence each other to produce sympathetic flares, with potentially stronger interactions occurring between closer regions, we would expect to see an overabundance of flares originating from nearby active regions at short waiting times compared to a simple Poisson distribution. The waiting time between two consecutive flares is defined as
\begin{equation}\label{eqn:w}
w = T_{\textrm{i+1}} - T_{\textrm{i}} \, ,
\end{equation}
where $T_{i}$ is the start time of the i'th flare occurring anywhere on the Sun. The angular distance between two flares is calculated using the Haversine formula, which determines the angular separation between two points on a sphere
\begin{equation}\label{eqn:haversine}
\lambda = 2\arcsin\,\left(\sqrt{\frac{1 - \cos\Delta\phi + \Bigl(1-\cos\Delta\theta\Bigl)\,\cos\phi_{1}\cos\phi_{2}}{2}}\,\right) \, ,
\end{equation}
\\
where $\lambda$ is the central angle, $\phi$ and $\theta$ are the longitude and latitude of a given point on the Sun, $\Delta\phi = \phi_{2} - \phi_{1}$ and $\Delta\theta = \theta_{2} - \theta_{1}$, with the indices 1 and 2 referring to the two consecutive flares. The longitude and latitude are expressed in the Carrington heliographic coordinate system to remove the effect of the solar rotation on the location of the flares. This ensures that no spatio-temporal correlation is introduced between multiple flares triggered from the same active region.  The various angles are shown in Figure \ref{fig:sphere}.

\begin{figure*}[h!]
    \centering \includegraphics[width = 2\columnwidth]{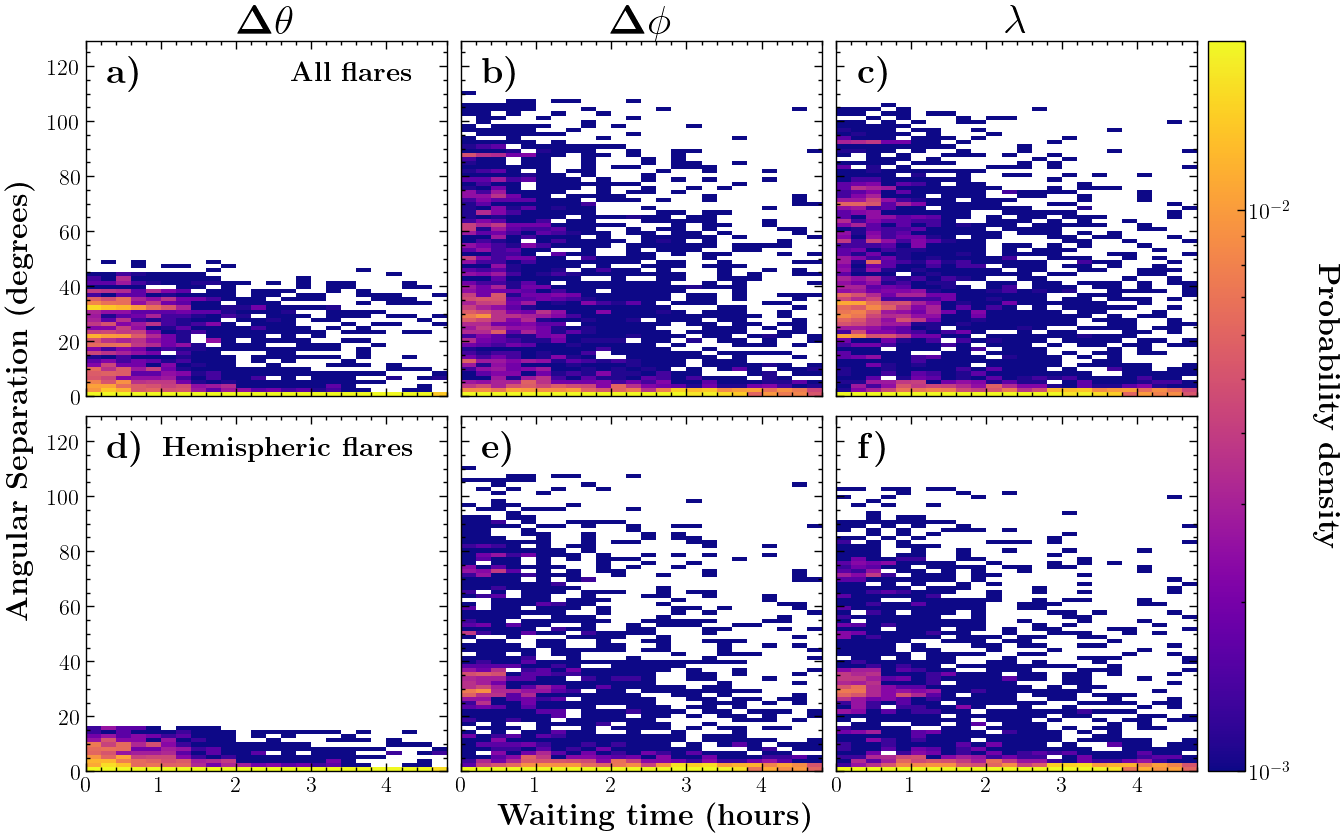}
    \caption{Distributions of angular separation in latitude (left), longitude (middle) and total angle (right) in degrees against waiting time in hours for flares observed by SDO/AIA between 2010-06-11 and 2019-01-26. In panels (a)-(c), all flares from both hemispheres are used while in panels (d)-(f), we first separate the flares from the northern and southern hemispheres and combine the distributions. The histograms have been normalized to show the probability densities.}
    \label{fig:sdo_2d}
\end{figure*}

\subsection{Statistical signature of sympathetic flares}

\begin{figure}[h!]
    \centering \includegraphics[width=\columnwidth]{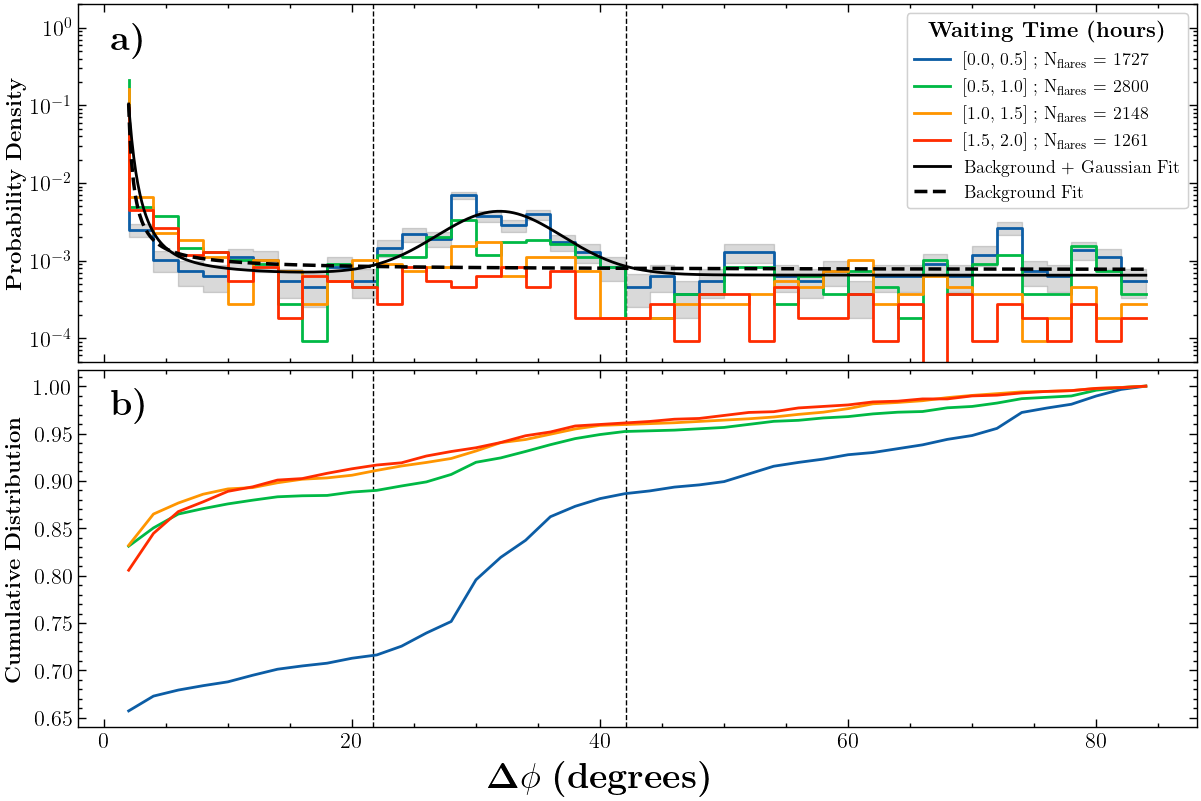}
    \caption{Longitudinal separation distribution between hemispheric flares from Figure \ref{fig:sdo_2d}(e) for different waiting time intervals (panel a). The solid and dashed black lines are fits of Equation \ref{eqn:fit}, with and without the Gaussian component respectively. The vertical dashed lines delimit the interval $\Delta\phi \in [\Delta\phi_{0}\pm2\sigma]$ from the Gaussian fit. Note the vertical logarithmic scale. Panel (b) shows the corresponding cumulative distribution function.}
    \label{fig:sdo_cdf_pdf}
\end{figure}

The values of $w$, $\,\lambda$, $\,\Delta\phi$ and $\Delta\theta$ are calculated for each pair of consecutive flares. In Figure \ref{fig:sdo_2d}, we show 2D histograms of the angular separation in latitude (left), longitude (middle) and total angle (right) against the waiting time for the flares observed by SDO/AIA. In the top panels (a-c), all flares from both hemispheres are used. In the bottom panels (d-f), we first measure the waiting time and angular separation between consecutive flares within each hemisphere separately, and then combine the distributions from the northern and southern hemispheres. This removes potential transequatorial sympathetic events, which are treated separately in Section \ref{sec:Transequatorial}. First, we observe a large number of events with very small angular separations across various waiting times. These represent pairs of consecutive flares triggered from the same active region, resulting in minimal separation. We exclude these events from our analysis, as we require sympathetic events to originate from distinct active regions, as historically defined. Then, from panels (a)-(c), we see an excess of events at short waiting times ($w \lesssim 1.5$ hour) between approximately $20^{\circ}$ to $40^{\circ}$ of separation. For $\Delta\theta$, this corresponds to consecutive flares triggered from different hemispheres within the latitudinal activity bands. In panel (d), this excess is no longer present because each hemisphere is considered individually, resulting in a smaller maximum separation in latitude between flares. However, the larger number of events at angular separations $\gtrsim 25^\circ$ persists for $\Delta\phi$ and $\lambda$ in panel (e) and (f) respectively, indicating that consecutive flares triggered from different active regions are more frequent within a given hemisphere compared to across hemispheres.
\\\\
To better show this excess of events, we present the $\Delta\phi$ distribution of consecutive hemispheric flares  in Figure \ref{fig:sdo_cdf_pdf}, with each histogram corresponding to a different waiting time interval. The histograms have been normalized to display the respective probability density functions (PDF). The bin error, $\epsilon =\sqrt{N}$, where $N$ is the number of events in that bin, is shown by the gray area. We see a clear excess of events around $\Delta\phi \approx 30^{\circ}$ that is increasingly pronounced for shorter waiting times. We model this behavior with a Gaussian function for the peak and an exponential decay function for the background distribution using
\begin{equation}\label{eqn:fit}
 \log\left(\textrm{PDF}\right) = A\,e^{-B/\Delta\phi^{2}} + C\,e^{-(\Delta\phi - \Delta\phi_{0})^{2} / 2\sigma^{2}} + D \, ,
\end{equation}
with A, B, C, D, $\Delta\phi_{0}$ and $\sigma$ being the free parameters. The fit using Equation \ref{eqn:fit}, overlaid on the blue histogram ($w \leq 0.5$ hour), is used to identify candidate sympathetic flaring events based on the following criteria: (1) $\Delta\phi \in$ [$\Delta\phi_{0}\pm2\sigma$] and (2) $w$ $\leq$ 1.5 hour, corresponding to flares well within the peak. A second fit, excluding the Gaussian component of Equation \ref{eqn:fit}, is also plotted in dashed to model the background distribution. By considering the ratio of these fits as a measure of the statistical significance of the signal, we find an average value of around 3 times more events within the peak. It is not possible to distinguish which events belong to the background and which are true sympathetic flares in the peak, but we do not expect background events to dominate. The cumulative distribution functions (CDF) in Figure \ref{fig:sdo_cdf_pdf}(b) illustrate that the excess of events is present regardless of the chosen histogram bins. The blue curve, corresponding to the shortest waiting time interval, shows a significant increase in the range $\Delta\phi \in [21^{\circ}, 42^{\circ}]$, clearly indicating the presence of a peak. This is slightly visible in the green curve as well (waiting time from 0.5 to 1 hour) but is insignificant for longer waiting times (red and oranges lines). A total of 581 candidate sympathetic flare pairs were identified, involving 922 unique flares, which account for approximately 7\% of all recorded flares. The reason the total number of unique flares is not exactly twice 581 pairs is that some flares are part of triplets, where the first and second flares are identified as sympathetic, as are the second and third for example.
\\
The peak around 30$^{\circ}$ is consistent with the findings of \citet{Fritzova-Svestkova1976}, who observed a statistical excess (3.4$\sigma$) of flares for pairs of active regions closer than 30$^{\circ}$ to each other for short waiting times ($w < 20$min). Similarly, \citet{Pearce1990} found a notable deviation from random flares for pairs of active regions separated by less than 35$^{\circ}$. Additionally, \citet{Schrijver2015} analyzed all SDO/AIA full-disk images associated with GOES flares of magnitude M5 or greater from May 1, 2010, to December 31, 2014. They observed an increase in the rate of subsequent flares within the first 4 hours, occurring at distances greater than 20$^{\circ}$ from the primary flare, with a significance level of 1.8$\sigma$. Therefore, the detection of a peak in this study at similar angular separations, based on a significantly larger number of flares compared to previous studies, provides strong evidence that the finding is both statistically significant and physically meaningful. In Appendix \ref{sec:AR}, we provide evidence that this peak does originate from the flares and not the underlying spatial distribution of active regions. Additionally, in Section \ref{sec:instruments}, we present results from other instruments (\textit{RHESSI} and Solar Orbiter/STIX) to further support this conclusion. 

\section{Properties of sympathetic flares} \label{properties}

In this section, we focus on the influence of the solar cycle on the presence of the peak, transequatorial flares as well as the correlation in magnitude and velocity propagation between consecutive flares.

\subsection{Variation over the solar cycle}\label{SolarCycle}

\begin{figure}[ht]
    \centering \includegraphics[width=\columnwidth]{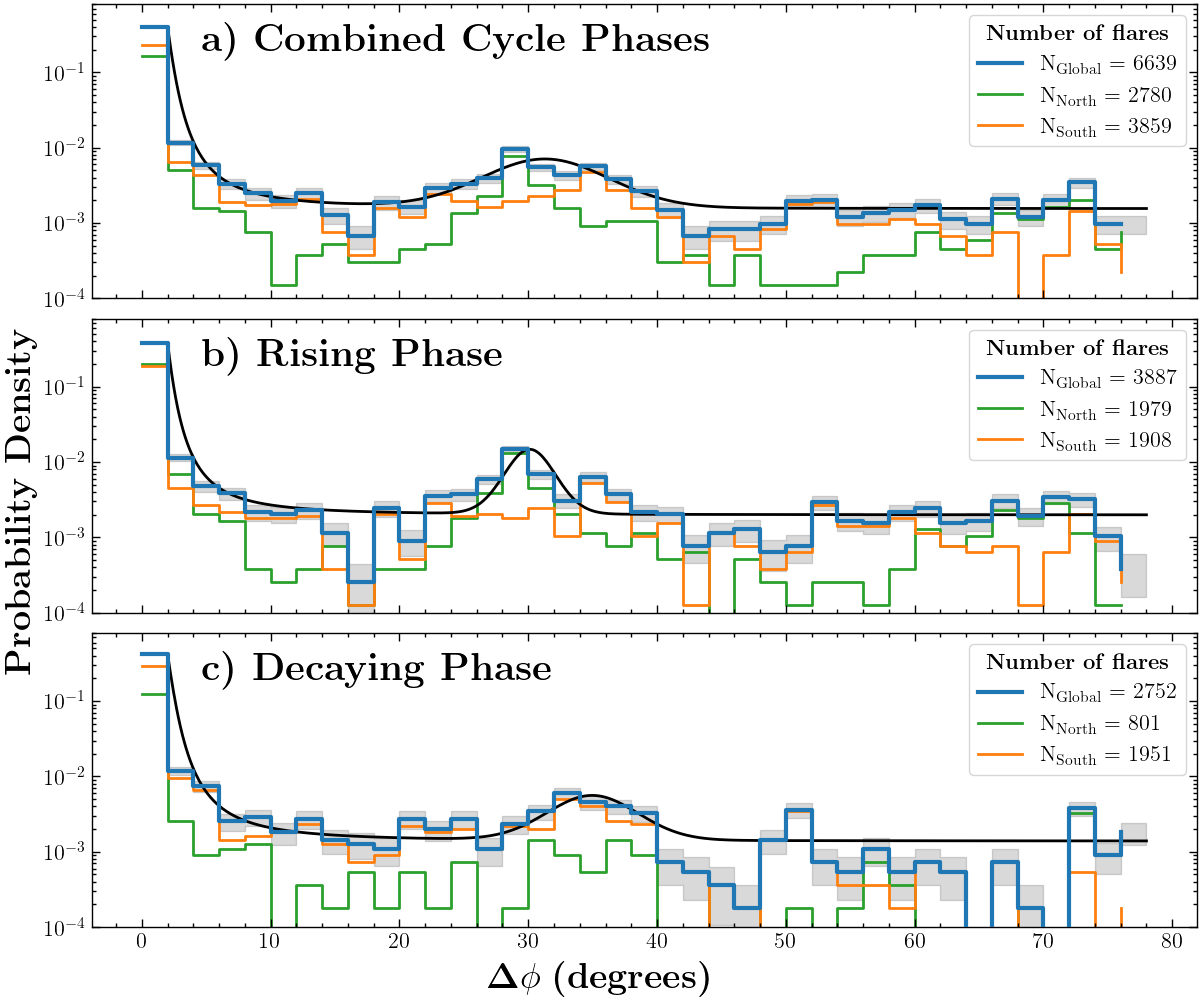}
    \caption{$\Delta\phi$ distribution of all hemispheric (blue), northern hemisphere (green) and southern hemisphere (orange) flare pairs with $w \leq 1.5$ hour. Panel (a) to (c) consider the whole cycle 24, the rising phase and the decaying phase, respectively. The time intervals are shown as vertical black dashed lines in Figure \ref{fig:SSN_TimeWindow}. The histograms are normalized so that the combined global distribution of both hemispheres integrates to unity. The black solid lines indicate fits of Equation \ref{eqn:fit} and the gray area indicates the $\sqrt{N}$ counting error.}
    \label{fig:SolarCycle}
\end{figure}

\begin{figure*}[h!]
    \centering 
    \includegraphics[width=2\columnwidth]{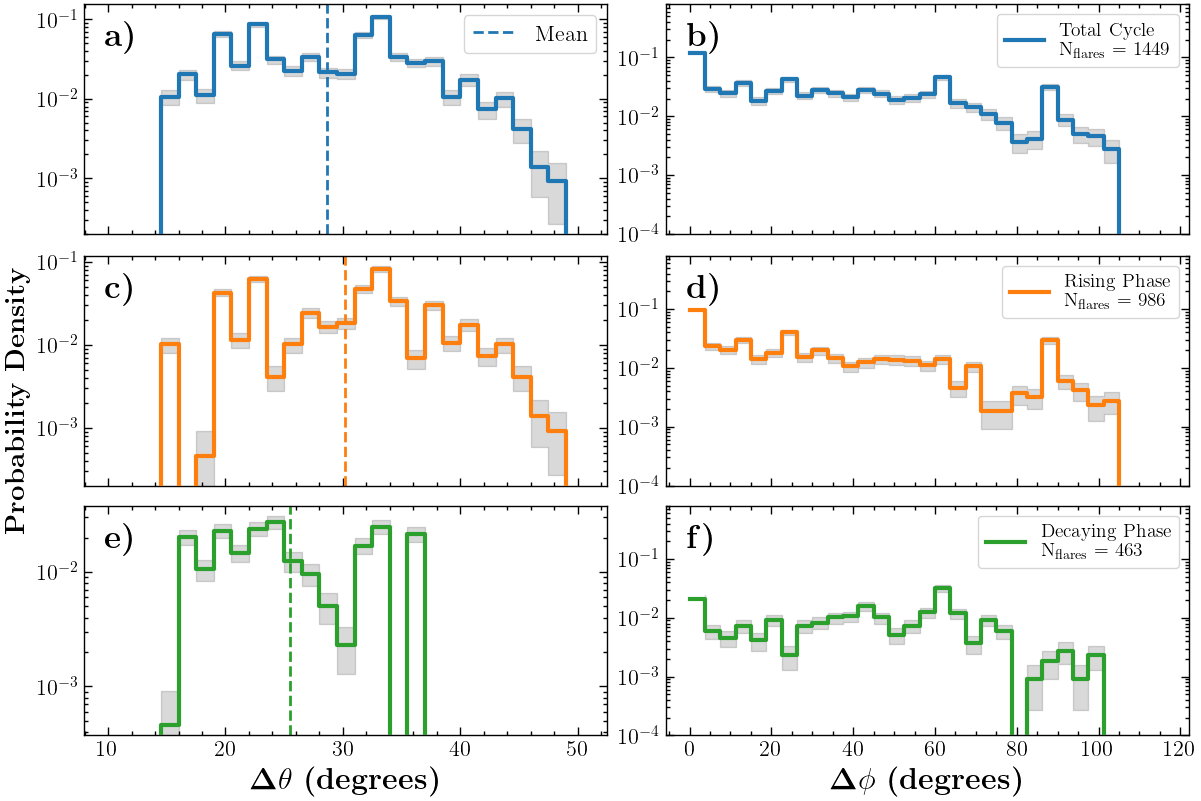}
    \caption{Angular separation in latitude (left) and longitude (right) between consecutive transequatorial flares observed by SDO/AIA with $w \leq 1.5$ hour. Panels (a)-(b) cover the full solar cycle 24, (c)-(d) the rising phase, and (e)-(f) the decaying phase. Phases match those in Figure \ref{fig:SolarCycle}, with vertical black dashed lines in Figure \ref{fig:SSN_TimeWindow}. Histograms are normalized so the total cycle (blue curve) integrates to unity. The gray area shows the $\sqrt{N}$ counting error, and dashed lines in (a) mark distribution means.}
    \label{fig:transequatorial}
\end{figure*}
\begin{figure}[h!]
    \centering \includegraphics[width=\columnwidth]{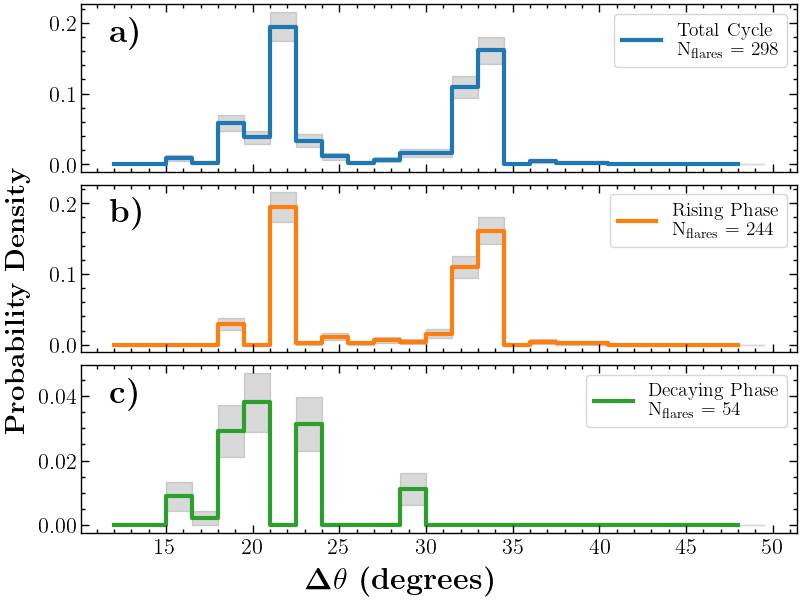}
    \caption{Same distributions as the left panels of Figure \ref{fig:transequatorial}, but only considering consecutive transequatorial flares with $\Delta\phi < 5^{\circ}$ and $w \leq 1.5$ hour, and with a vertical linear scale. The gray area indicates the counting error $\sqrt{N}$.}
    \label{fig:transequatorial_mask}
\end{figure}

\noindent The distributions shown in Figure \ref{fig:sdo_cdf_pdf} are based on the SDO/AIA hemispheric flares covering nearly the entire solar cycle 24. However, a similar analysis can be conducted by separately examining the rising and decaying phases of the cycle as well as the northern and southern hemispheric flares. We present in Figure \ref{fig:SolarCycle} the distribution in $\Delta\phi$ for almost the entire cycle 24 (panel a; 2010-06-01 to 2018-03-01), the rising phase (panel b; 2010-06-01 to 2014-03-31) and the decaying phase (panel c; 2014-04-01 to 2018-03-01). The vertical black dashed lines in Figure \ref{fig:SSN_TimeWindow} indicate these time intervals, which are consistently applied to all subsequent analyses examining different phases of the cycle, unless specified otherwise. These time intervals were selected to cover the overlap between SDO and \textit{RHESSI}, while also dividing the solar cycle at the point where the sunspot number is maximal, into rising and decaying phases. This division results in a comparable number of flares between the rising and decaying phases, as well as an almost equal number of flares between hemispheres during the rising phase. We only consider flare pairs with $w \leq 1.5$ hour, but do not divide them into waiting time bins to have enough events when splitting the data by hemisphere and solar cycle phase. In each panel, we plot the distribution for all hemispheric flare pairs (blue), northern hemisphere flare pairs (green) and southern hemisphere flare pairs (orange) and show the corresponding probability densities, normalized so that the global distribution (blue curve) sums to unity. The black solid lines represent fits using Equation \ref{eqn:fit}, which, while not entirely accurate for large $\Delta\phi$, effectively model both the peak and the smaller values of $\Delta\phi$. In panel (a), when looking at both hemispheres separately, we see that the northern hemisphere (in green) has its peak slightly to the left while the southern hemisphere (in orange) has its peak to the right. This behavior is even more apparent when looking at the different phases in the solar cycle. For the rising phase in panel (b), the global hemispheric peak seems to be slightly dominated by the contribution from the northern hemisphere flares (with $\Delta\phi_{0} = 30^{\circ} \pm 3^{\circ}$). In contrast, the southern hemisphere flares dominate the decaying phase of the cycle in panel (c), with a broader peak ($\Delta\phi_{0} = 34^{\circ} \pm 7^{\circ}$). This is consistent with the asymmetry of solar cycle 24, where the southern hemisphere peaked in activity significantly later than the northern hemisphere \citep[e.g.][]{Veronig2021}, resulting in far fewer northern hemisphere flares during the decaying phase. However, the observed shift in the peak's position is not due to cycle hemispheric asymmetry, as the rising phase includes the activity peak of both hemispheres. These results are not significantly impacted by splitting the cycle at a slightly different point in time.

\subsection{Transequatorial flares}\label{sec:Transequatorial}

The previous analyses initially filtered consecutive flares from the northern and southern hemispheres separately and then combined the resulting distributions. However, this method excluded potential transequatorial sympathetic flares, which are the focus of this section. In Figure \ref{fig:transequatorial}, we present the angular distribution of latitudinal (left) and longitudinal (right) separations of consecutive flares triggered from two separate hemispheres with $w \leq 1.5$ hour. Panel (a) and (b) include flares covering almost the entire solar cycle 24 , panel (c) and (d) only include the rising phase, and panel (e) and (f) only the decaying phase. Panel (a) shows that the distribution of $\Delta\theta$ for cycle 24 is approximately flat, which results from a gradual shift from larger latitudinal separation during the rising phase (panel c) to smaller separations during the decaying phase (panel e). This is to be expected, as the locations where active regions emerge migrate towards the equator from the minimum to the maximum of the cycle, as illustrated by the standard butterfly time-latitude diagram (e.g. \citealt{hathaway2015}). In contrast, the $\Delta\phi$ distributions appear to be independent of the solar cycle phase, showing a roughly flat distribution over a large range of separations. However, there is a clear excess of flares for $\Delta\phi \lesssim 5^{\circ}$ when considering that the vertical scale is logarithmic. These events are distinct from the excess in Figure \ref{fig:sdo_cdf_pdf}(a), as we now focus on consecutive transequatorial flares that cannot be triggered by the same active region.
\\\\
In Figure \ref{fig:transequatorial_mask}, we analyze the $\Delta\theta$ distribution of consecutive transequatorial flares with $\Delta\phi < 5^{\circ}$. Panel (a) shows a large deficit of events around $\Delta\theta \approx 25^{\circ} - 30^{\circ}$ despite the small sample size, in contrast to the relatively flat distribution in Figure \ref{fig:transequatorial}(a), where a much smaller deficit was visible. This indicates that as the cycle progresses, there is an epoch between the rising and decaying phases when transequatorial flares at similar longitudes become significantly less frequent. Interestingly, this occurs at an angular separation in latitude ($\Delta\theta \approx 30^{\circ}$) that is comparable to the longitudinal peak of hemispheric flares ($\Delta\phi \approx 30^{\circ}$), which is quite surprising. This suggests a kind of negative correlation, where the trigger of a flare in one hemisphere might reduce the probability of a flare in the opposite hemisphere. To the best of our knowledge, this phenomenon has not been observed or described in previous studies. We refer to it as \textit{unsympathetic flares}.

\subsection{No correlation in the magnitude of two consecutive flares}
\begin{figure}[ht]
    \centering \includegraphics[width=\columnwidth]{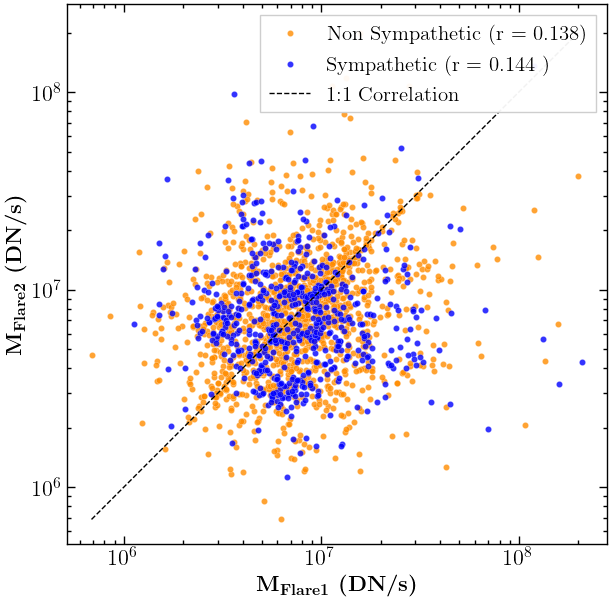}
    \caption{131 $\AA$ SDO/AIA peak magnitude of the second flare is plotted against that of the first flare for each pair of consecutive non-sympathetic (orange) and sympathetic (blue) flares. The peak magnitude is expressed in Data Number per second, representing raw detector output not calibrated to irradiance units. The black dashed line corresponds to the 1:1 correlation line.
    }
    \label{fig:magnitude}
\end{figure}

From the SDO/AIA flare list, we can use the peak magnitude at a given wavelength (in Data Numbers per second, i.e. raw detector output not calibrated to irradiance units) as a proxy for its energy when available. We use the magnitude of the 131$\AA$ AIA channel, since \citet{vanderSande2022} found that it correlates best with GOES X-ray magnitudes. In Figure \ref{fig:magnitude}, for each pair of consecutive flares, we show the peak magnitude of the second flare against the peak magnitude of the first flare, with non-sympathetic flares shown in orange and sympathetic flares shown in blue. For both types of flares, we do not observe a clear correlation between the peak magnitudes, as the data points scatter up to one order of magnitude away from the 1:1 correlation line. From Figure \ref{fig:magnitude}, we can conclude that while sympathetic flares are believed to be connected, this does not necessarily indicate a strong coupling between the energy of the first flare and that of the second. 
Put differently, it means that in the SDO/AIA sample, we find sympathetic flares can occur such that a small flare helps triggering a large flare, or the opposite, without any preferred situation. This is contrary to the findings of \citet{Mawad2022}, who reported that approximately 78\% of their candidate sympathetic flares were of the twin type, meaning the secondary flare had the same GOES class as the primary flare. 

\subsection{Propagation velocity}
\begin{figure}[h!]
    \centering \includegraphics[width=\columnwidth]{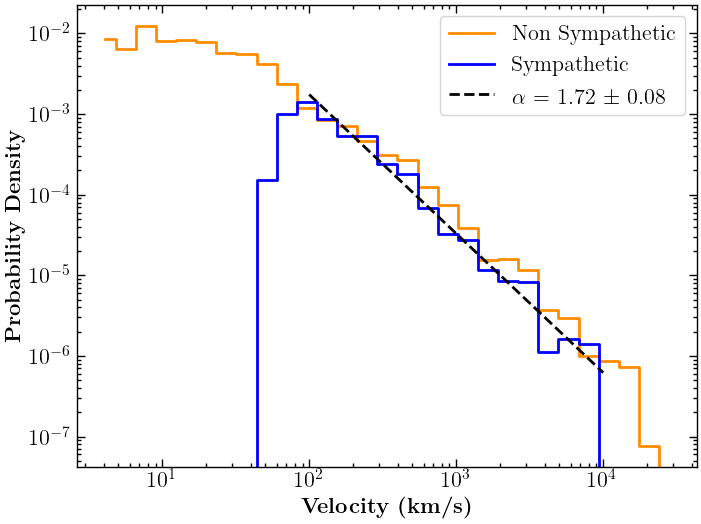}
    \caption{Distribution of propagation velocities for a signal traveling the angular distance $\lambda$ over the waiting time $w$ between consecutive non sympathetic (orange) and sympathetic (blue) flares. The black dashed line indicates a power-law fit. We only include non-sympathetic flares with $w < 6$ hours and separated by $\lambda > 5^{\circ}$ to exclude flares triggered by the same active region.}
    \label{fig:velocity}
\end{figure}

Assuming a causal link between two sympathetic events, we calculate the propagation velocity needed for a signal to travel the angular distance $\lambda$ across the solar surface between the flares within a time $w$. This gives a lower bound for the propagation velocity, as $\lambda$ corresponds to the shortest separation between the flares. Additionally, we do not account for a specific coronal flaring height, since it is small compared to the solar radius and is unlikely to significantly impact the results. No assumptions are made regarding the specific physical mechanism driving the interaction. The velocity distribution is shown in Figure \ref{fig:velocity}, with the blue and orange curves for sympathetic and non-sympathetic flares respectively. For the non-sympathetic flares, we only include consecutive flares separated by $\lambda > 5^{\circ}$ and $w < 6$ hours to exclude flares triggered by the same active region, despite the fact that a propagation velocity is technically not relevant for non-sympathetic flares. For both distributions, we observe a power-law behavior ($f(v) = f_{0}v^{-\alpha}$ with $v$ the velocity) with a slope $\alpha = 1.72 \pm 0.08$ for $v \gtrsim 100$ km/s, shown by the black dashed line. In the case of candidate sympathetic flares, there is a clear cutoff of the distribution below $v \approx 80$ km/s. This value can put constraints regarding the possible interaction mechanism for sympathetic flares, which will be discussed more in depth in Section \ref{sec:interpretation}. A cutoff in the distribution is expected, as sympathetic flares are separated by at least $\sim20^{\circ}$ and occur within $w \leq 1.5$ hours, according to our selection criteria. However, the cutoff speed for a minimum distance of 20$^{\circ}$ and a maximum time of 1.5 hours is 45 km/s. Thus, the observed cutoff of 80 km/s is significantly higher than the minimal expected value. Although there is an excess of hemispheric flares with $w \leq 1.5$ hour and $\Delta\phi \approx 30^{\circ}$, there is not a distinctive characteristic velocity for sympathetic flares, as both distributions are similar. This likely reflects the absence of a characteristic scale in the distribution of active regions, as shown in Appendix \ref{sec:AR}.

\begin{figure*}[h]
    \centering 
    \includegraphics[width = 2\columnwidth]{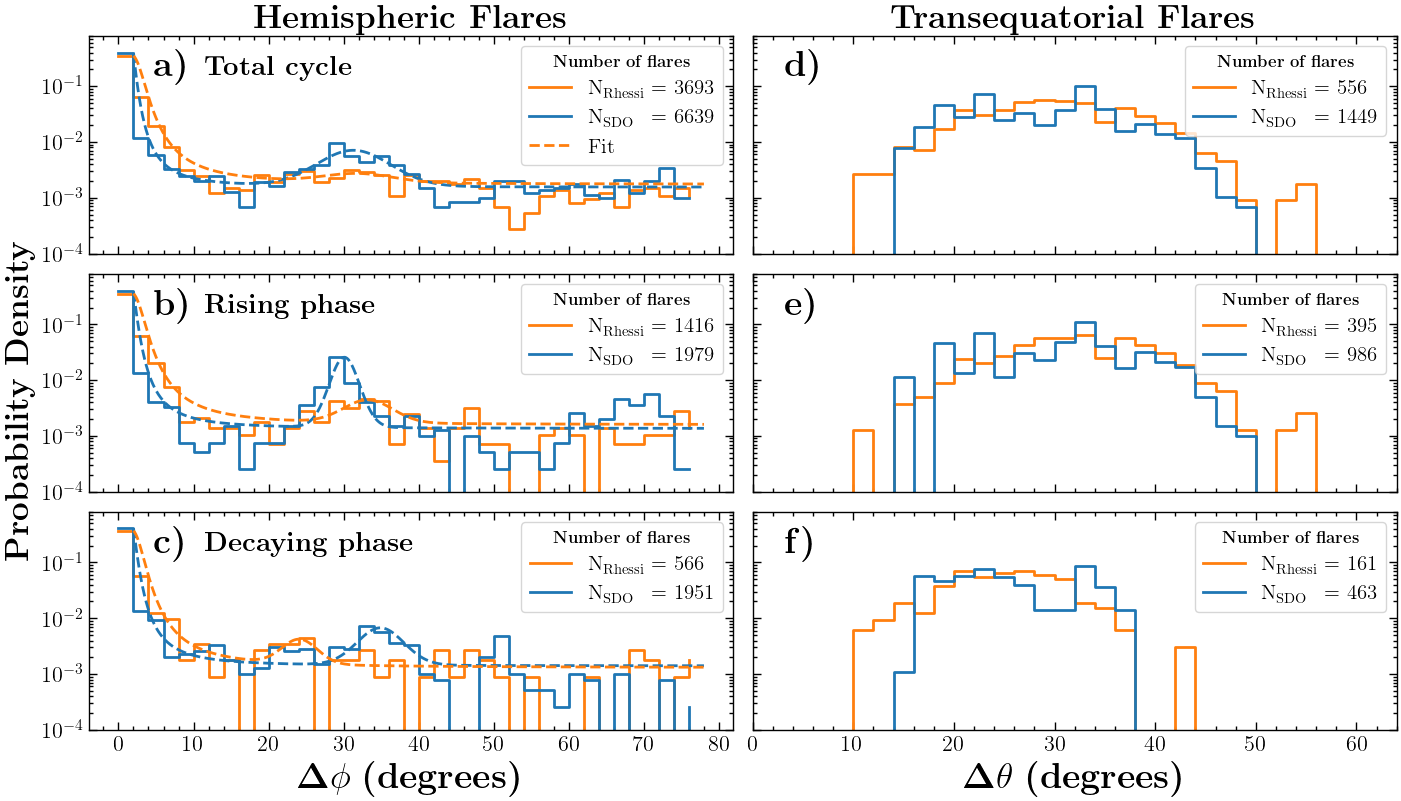}
    \caption{Distribution of $\Delta\phi$ for hemispheric flares (panels a-c) and $\Delta\theta$ for transequatorial flares (panels d-f) with $w \leq 1.5$ hours, comparing \textit{RHESSI} (orange) and SDO/AIA (blue). The first, second and third row correspond to the total cycle, the rising phase and the decaying phase, respectively. The fits of Equation \ref{eqn:fit} are shown as dashed lines. Northern hemisphere and southern hemisphere flares are used in panel (b) and (c) respectively to maximize the signal (see Figure \ref{fig:SolarCycle} for hemispheric asymmetry).}
    \label{fig:SDO_RHESSI}
\end{figure*}

\begin{figure}[h]
    \centering 
    \includegraphics[width = \columnwidth]{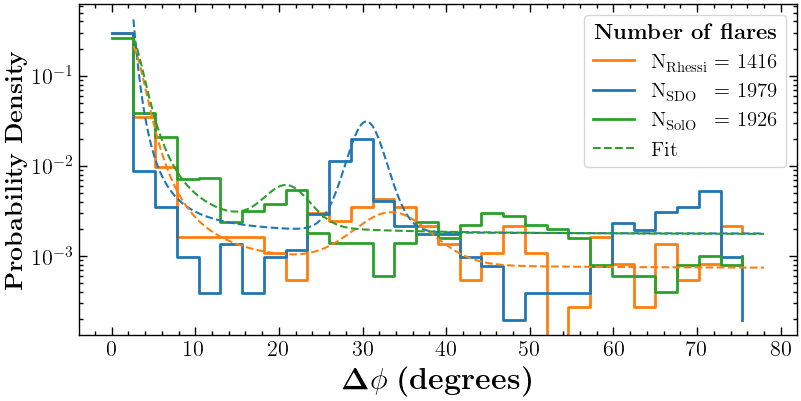}
    \caption{Same distributions as  Figure \ref{fig:SDO_RHESSI}(b), with the addition of flares observed by Solar Orbiter/STIX (green curve). The fits of Equation \ref{eqn:fit} are shown as dashed lines.}
    \label{fig:SDO_RHESSI_STIX}
\end{figure}

\section{Comparison between instruments}\label{sec:instruments}

\subsection{SDO and \textit{RHESSI} during cycle 24}\label{sec:sdo_rhessi}
The large overlap in observations between SDO/AIA and \textit{RHESSI}, as shown in Figure \ref{fig:SSN_TimeWindow}, allows to compare both instruments for the analyses conducted in Sections \ref{SolarCycle} and \ref{sec:Transequatorial}. In Figure \ref{fig:SDO_RHESSI}, we present the $\Delta\phi$ distribution of hemispheric flares (left) and the $\Delta\theta$ distribution of transequatorial flares (right) with $w \leq 1.5$ hours observed by SDO/AIA (blue) and \textit{RHESSI} (orange). The first, second and third row correspond to the total cycle, the rising phase and the decaying phase respectively. Since flares in the northern hemisphere are more prevalent during the rising phase of the solar cycle, and those in the southern hemisphere dominate during the decaying phase (see Figure \ref{fig:SolarCycle}), we consider only northern hemisphere flares in panel (b) and southern hemisphere flares in panel (c). This filtering makes the signal clearer, allowing for a more effective comparison between the instruments. The fits of Equation \ref{eqn:fit} are shown as dashed lines. For hemispheric flares during the rising phase (panel b), \textit{RHESSI} generally agrees with SDO/AIA, showing a similar peak, although slightly broader ($\Delta\phi_{0} = 33^{\circ} \pm 7^{\circ}$). In the decaying phase, a small peak ($\Delta\phi_{0} = 24^{\circ} \pm 5^{\circ}$) is fitted for the \textit{RHESSI} distribution, but the smaller number of flares makes this fit neither representative nor statistically significant. As for the transequatorial flares, the two distributions are quite similar, with a slightly larger $\Delta\theta$ span for \textit{RHESSI} compared to SDO/AIA.

\subsection{Rising phase analysis for SDO/AIA, \textit{RHESSI} \& Solar Orbiter/STIX}\label{sec:sdo_rhessi_stix}

As shown in Figure \ref{fig:SolarCycle}, the $\Delta\phi$ position of the peak, identified as candidate sympathetic flares, varies depending on the solar cycle phase. Since Solar Orbiter/STIX observations are limited to the rising phase of cycle 25, we only use SDO/AIA and \textit{RHESSI} flares observed during the rising phase of cycle 24 for the comparison. Figure \ref{fig:SDO_RHESSI_STIX} shows the $\Delta\phi$ distribution of hemispheric flares with $w \leq 1.5$ hours from \textit{RHESSI} (orange), SDO/AIA (blue), and Solar Orbiter/STIX (green), which shows a similar peak although at a smaller angular separation ($\Delta\phi_{0} = 22^{\circ} \pm 6^{\circ}$) and with a lower amplitude. The fits of Equation \ref{eqn:fit} are shown as dashed lines. The locations of flares observed by STIX are currently determined using a sub-collimator known as the Coarse Flare Locator (CFL; \citealt{Krucker2020}). A recently developed neural network offers enhanced accuracy in estimating flare coordinates \citep{massa2024} compared to the current method, which could potentially improve our analysis. The detection of a peak in the Solar Orbiter/STIX data indicates that the position of the peak may vary between solar cycles, as the alignment between SDO and \textit{RHESSI}, both of which observed during cycle 24, is better compared to Solar Orbiter, which observed during cycle 25. Nonetheless, despite some variations between instruments, data from three different instruments with observations across multiple solar cycles consistently indicate the presence of a peak in nearby hemispheric flares around $\Delta\phi \approx 20^{\circ}-30^{\circ}$ at short waiting times $w \leq 1.5$ hour. This confirms and refines the initial findings of \citet{Fritzova-Svestkova1976}, \citet{Pearce1990} and \citet{Schrijver2015} and provides good evidence for the statistical existence of sympathetic flares.

\section{List of candidate sympathetic flares}\label{sec:list_candidate}
In Table \ref{table:sympathetic_flares}, we present a small representative sample of flare pairs identified as sympathetic events from the SDO/AIA flare list. The table lists the start, peak and end times in the 131 $\AA$ filter, Heliographic Carrington coordinates, and 131 $\AA$ magnitude for each flare in the sympathetic pairs, with a horizontal line separating each pair of events. The table containing all the candidate sympathetic events for each instrument is available online.\footnote{\url{https://github.com/lsguite/sympathetic_flares_catalogs.git}}

\section{Discussion and conclusions} \label{sec:interpretation}
We analyzed a large number of flares observed by SDO/AIA, \textit{RHESSI} and Solar Orbiter/STIX in the context of a statistical study of sympathetic flares, leading to several key findings. First, a statistically significant peak in the longitudinal distribution of consecutive hemispheric SDO/AIA flares is observed at $\Delta\phi = 31^{\circ} \pm 10^{\circ}$ (see Figure \ref{fig:sdo_cdf_pdf}). This peak was specifically observed for consecutive flares separated by a short waiting time ($w \lesssim$ 1.5 hour). A similar peak was observed by the two other instruments (\textit{RHESSI} and Solar Orbiter/STIX), although at slightly different $\Delta\phi$ values and with varying intensities (see Figure \ref{fig:SDO_RHESSI_STIX}). The number of identified candidate sympathetic flares correspond to 7\%, 3.5\% and 3.9\% for SDO/AIA, \textit{RHESSI} and Solar Orbiter/STIX respectively, giving an average of approximately 5\% across all three instruments. Second, the angular separation of the peak varies throughout the solar cycle (24 for SDO/AIA), with $\Delta\phi = 30^{\circ} \pm 3^{\circ}$ during the rising phase and $\Delta\phi = 34^{\circ} \pm 7^{\circ}$ in the decaying phase (see Figure \ref{fig:SolarCycle}). Third, when considering consecutive transequatorial flares with $\Delta\phi \lesssim 5^{\circ}$, a strong deficit is observed around $\Delta\theta \approx 25^{\circ}$ (see Figure \ref{fig:transequatorial_mask}). Interestingly, this roughly coincides with the location of the $\Delta\phi$ peak observed for hemispheric flares. \\
\begin{table*}[h!]
    \centering
    \footnotesize 
    \caption{Sample of SDO/AIA candidate sympathetic flares \label{table:sympathetic_flares}}
    \begin{tabular}{ccccccc} 
        \hline
        \hline
        \vspace{-0.25cm} \\ Flare ID & 
        Start time (UTC) & 
        Peak time (UTC) & 
        End time (UTC) & 
        Longitude ($^{\circ}$) & 
        Latitude ($^{\circ}$) & 
        131$\AA$ Magnitude ($10^{6}$ DN/s) \vspace{0.05cm} \\
        \hline
        \vspace{-0.25cm} \\ 8116  & 2011-03-09 04:39:00 & 2011-03-09 04:43:00 & 2011-03-09 05:18:00 & 90.541669  & 8.752714   & 4.721330 \vspace{0.05cm} \\
        8232  & 2011-03-09 04:43:00 & 2011-03-09 04:51:00 & 2011-03-09 05:43:00 & 62.333471  & 19.780159  & 3.694895 \vspace{0.05cm} \\
        \hline
        \vspace{-0.25cm} \\ 14940 & 2011-07-29 11:59:00 & 2011-07-29 12:03:00 & 2011-07-29 12:13:00 & 358.433534 & 19.639778  & 4.911795 \vspace{0.05cm}  \\
        14978 & 2011-07-29 12:12:00 & 2011-07-29 12:17:00 & 2011-07-29 12:42:00 & 327.583848 & 16.203819  & 10.70984 \vspace{0.05cm}  \\
        \hline
        \vspace{-0.25cm} \\ 4573  & 2013-08-11 16:25:00 & 2013-08-11 16:43:00 & 2013-08-11 17:21:00 & 214.821658 & -7.024604  & 3.201248 \vspace{0.05cm}  \\
        4545  & 2013-08-11 16:36:00 & 2013-08-11 16:48:00 & 2013-08-11 16:59:00 & 240.638774 & -21.923117 & 5.987593 \vspace{0.05cm}  \\
        \hline
        \vspace{-0.25cm} \\ 10388 & 2014-12-17 02:22:00 & 2014-12-17 02:32:00 & 2014-12-17 02:41:00 & 216.061398 & -9.972204  & 7.837965 \vspace{0.05cm}  \\
        10278 & 2014-12-17 03:10:00 & 2014-12-17 03:19:00 & 2014-12-17 03:26:00 & 249.175753 & -16.107000 & 24.621448 \vspace{0.05cm} \\
        \hline
        \vspace{-0.25cm} \\ 13653 & 2016-02-10 14:39:00 & 2016-02-10 14:42:00 & 2016-02-10 14:45:00 & 128.475396 & 12.318803  & 10.61923 \vspace{0.05cm}  \\
        13654 & 2016-02-10 15:13:00 & 2016-02-10 15:22:00 & 2016-02-10 15:48:00 & 91.091178  & 11.124230  & 13.00206 \vspace{0.05cm} \\
        \hline
    \end{tabular}
    \tablefoot{The values are sourced from \cite{vanderSande2022} for a sample of the candidate sympathetic flares. The flare coordinates are provided in the Heliographic Carrington system, whereas in the original list they were expressed in the Heliographic Stonyhurst system. The start, peak, and end times correspond to the flaring interval in the 131 $\AA$ SDO/AIA channel. The flare ID corresponds to the row number in the original \cite{vanderSande2022} flare list. The magnitude is given in units of Data Numbers per second, i.e., raw detector output not calibrated to irradiance units. The link to the online table of all candidate sympathetic events for each instrument is provided at the end of Section \ref{sec:list_candidate}.}
\end{table*}

\begin{figure}[h!]
    \centering
\includegraphics[width=\columnwidth]{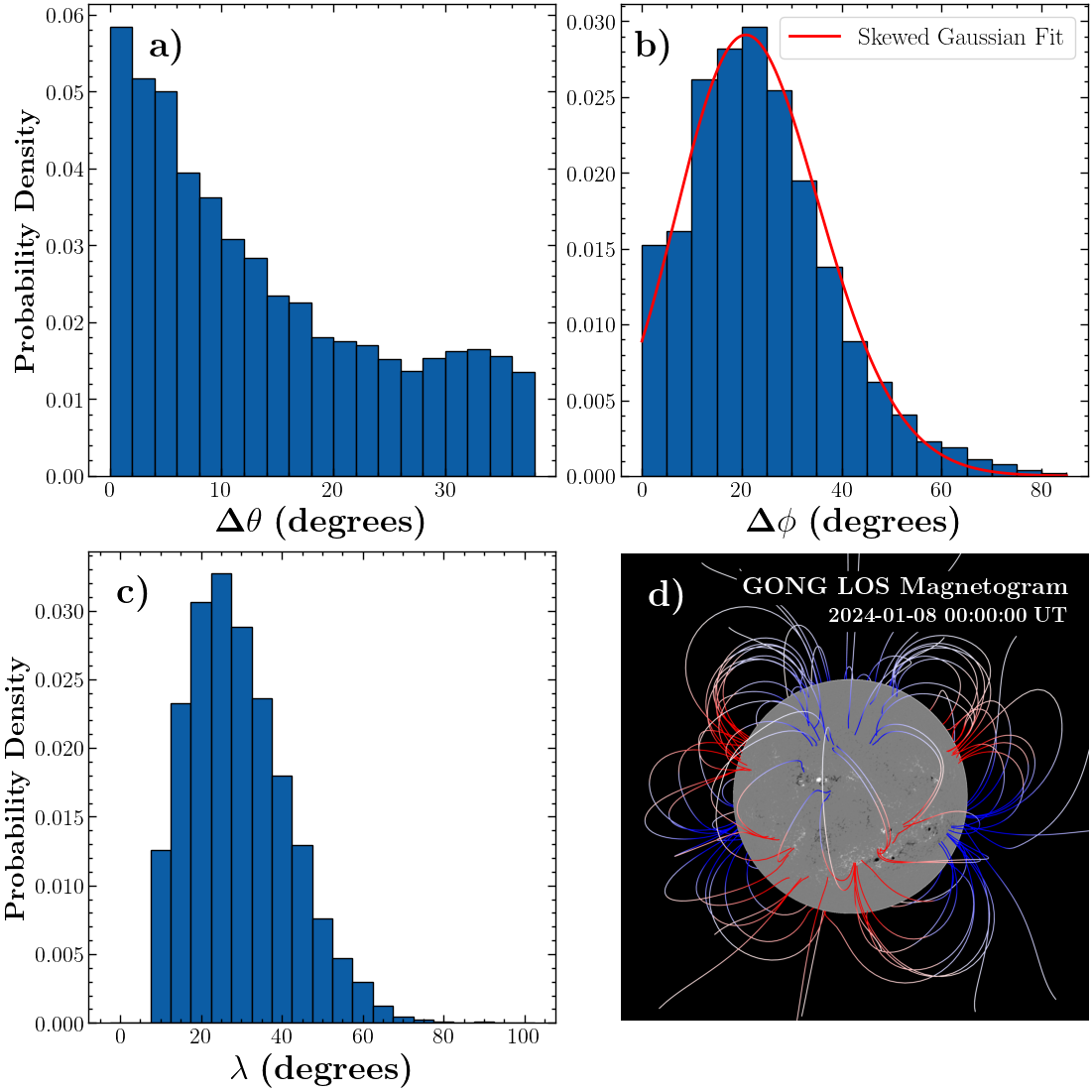}
    \caption{Distribution of the angular separation in latitude (a), longitude (b) and total angle (c) between the footpoints of closed magnetic field lines. We consider field lines whose footpoints are on the same hemisphere and separated by $\lambda \geq 10^{\circ}$ to exclude field lines closing in a single active region. The field lines were generated by PFSS extrapolations made using the \texttt{pfsspy} code from \citet{Stansby2020}, with the source surface radius at $r_{SS} = 2.5$R$_{\odot}$ and using GONG magnetograms between 2010-06-16 and 2018-02-23 as photospheric boundary conditions. The time intervals approximately correspond to those indicated by the vertical black dashed lines in Figure \ref{fig:SSN_TimeWindow}. In panel (d), we show an example of a PFSS extrapolation using a GONG magnetogram on 2024-06-10 in JHelioviewer \citep{Muller2017}. In panel (c), the red line is a fit of a skewed Gaussian function: $y = \textrm{exp}\{-(x-\mu)^{2}/2\sigma^{2}\}\cdot\{1 + \textrm{erf}(\alpha(x-\mu)/\sqrt{2}\sigma)\} \, / \sqrt{2\pi}\sigma$}
    \label{fig:pfss_distribution}.
\end{figure}

\noindent This suggests that 30 degrees may be a characteristic length scale, either influencing the triggering of additional sympathetic flares (hemispheric) or their suppression (transequatorial). While one might speculate that this deficit compensates for the excess in the $\Delta\phi$ distribution of hemispheric flares, there is no requirement for flare events to be evenly distributed in space. In other words, the hemispheric $\Delta\phi$ and the transequatorial $\Delta\theta$ distributions are entirely independent of each other. Moreover, the results of Figure \ref{fig:hek_ar_transequatorial} indicate that this deficit is not present in the underlying spatial distribution of active regions. The asymmetry of solar cycle 24, in which the northern hemisphere peaks in activity before the southern hemisphere \citep[e.g.][]{Veronig2021}, may lead to variations in the number of consecutive transequatorial flares throughout the cycle. However, this should not depend on the specific range of $\Delta\phi$ considered. Overall, we are currently unable to offer a definitive explanation for the observed signal. We propose the term \textit{unsympathetic flares} to describe this phenomenon, where the occurrence of a flare in one active region may reduce the likelihood of a subsequent flare in a region located at the same longitude on the opposite hemisphere. 
\\\\
For the identified candidate sympathetic flares, the length scale of $\Delta\phi \approx 30^{\circ}$ and waiting times of $w \leq 1.5$ hour corresponds to a distribution of propagation velocities with a peak at approximately $v \approx 80$ km/s and a maximum value of $v \approx 10^{4}$ km/s (see Figure \ref{fig:velocity}). These velocities can be compared to the characteristic speeds of known physical phenomena, such as EUV waves, Alfvén waves or mass transfer for example. In the case of coronal EUV waves, \citet{Warmuth2011} carried out a statistical analysis using a sample of 176 events and identified three distinct velocity groups: $v \gtrsim 320$ km/s, $v \approx 170-320$ km/s and $v \lesssim 130$ km/s. Additionally, \citet{Cunha2018} reported both fast ($v \approx 560$ km/s) and slow ($v \approx 250$ km/s) EUV wave fronts based on SDO/AIA images. These velocities could explain the sympathetic flares from the lower part of the propagation velocity distribution. As for Alfvén waves, \citet{Aschwanden1999a} analyzed a set of 30 loops observed by SoHO/EIT \citep{Poland1990, Delaboudiniere1995} to investigate the height dependence of various physical parameters in the corona, including the Alfvén velocity. At the loop footpoints, the Alfvén velocity ranged from $v_{A} = 2000$ to 6000 km/s, while at the loop tops (h = 100 Mm), it decreased to between $v_{A} = 500$ and 1000 km/s. Furthermore, \citet{Warmuth2005} developed an analytical model for the Alfvén speed in the solar corona, calculating a value of $v_{A} \geq 2000$ km/s at a height of $10$ Mm above an active region. The 131 $\AA$ emission, probed by one of the SDO/AIA channel, typically forms at a height of about 1 Mm \citep{Sanjay2024}, whereas \textit{RHESSI} flares are most often observed at approximately 6 Mm \citep{Christe2011}. Thus, these Alfvén velocity estimates may be applicable to the flares in this study and could account for the upper range of the sympathetic flare distribution. Finally, \citet{Wang2001} conducted a multi-instrument study of a sympathetic flaring event on February 17th, 2000, concluding that the second eruption was triggered by a mass surge propagating at approximately 80 km/s, following the first filament eruption and its associated flare. Remarkably, this velocity 
matches the peak value of the propagation velocity distribution for the candidate sympathetic flares identified in this study. In summary, the propagation velocities mentioned above all fall within the range shown in Figure \ref{fig:velocity}, suggesting they are plausible mechanisms for mediating interactions between separate flaring active regions, based on the observed distances and waiting times between candidate sympathetic flares. However, the fastest velocities in the distribution seem too high for mass surges, implying that only a subset of candidate sympathetic flares could be triggered by this mechanism. Nonetheless, our analysis cannot determine which mechanism is most likely to dominate in sympathetic eruptions
\\\\
Until now, no explanation has been provided for why an angular separation of approximately 30$^{\circ}$ might favor sympathetic flares. To address this, we estimate the mean magnetic connectivity on the Sun, i.e. the average angular separation between the footpoints of magnetic field lines. For each Carrington rotation between 2010-06-16 and 2018-02-23, we perform a PFSS extrapolation using the \texttt{pfsspy} code from \citet{Stansby2020}, setting the source surface radius to $r_{SS} = 2.5$R$_{\odot}$. GONG magnetograms are used to define the photospheric boundary conditions of the radial magnetic field. The field lines are traced from seed points distributed uniformly, in spherical coordinates, with 120 points spanning 360$^{\circ}$ in longitude and 30 points spanning $\pm 45^{\circ}$ in latitude. This results in a total of more than 370,000 field lines traced when combined over the Carrington rotations mentioned above. For every closed field line in the PFSS extrapolations, we then calculate the total angular separation, along with the longitudinal and latitudinal separations between the field line footpoints. The corresponding distributions are shown in Figure \ref{fig:pfss_distribution}, with the separation in latitude (a), longitude (b), and total angular separation (c). We consider only field lines with footpoints separated by $\lambda \geq 10^{\circ}$ to exclude those confined within a single active region. Of course, sympathetic eruptions may (and do) originate within the same active regions or within close proximity ($\leq10^{\circ}$). These are usually referred to as homologous flares rather than sympathetic \citep[e.g.][]{Martres1989}. While such flares are present in our data (see Figure \ref{fig:sdo_cdf_pdf} for numerous events within the bin $\Delta\phi \lesssim 10^{\circ}$), this study focuses exclusively on flare pairs originating from distinct active regions. Panel (d) shows a visual example of a PFSS extrapolation using a GONG magnetogram on 2024-06-10 in JHelioviewer \citep{Muller2017}. The $\Delta\phi$ distribution in panel (b) exhibits a distinct peak, whose shape can be estimated by fitting a skewed Gaussian function. From the fit, we estimate the peak location ($\phi_{\textrm{peak}}$) and the separations corresponding to half the peak's probability density function value ($\phi_{1}$ and $\phi_{2}$). The upper uncertainty is given by $\sigma_{+} = \phi_{2} - \phi_{\textrm{peak}}$, and the lower uncertainty by $\sigma_{-} = \phi_{\textrm{peak}} - \phi_{1}$. This results in a peak separation of $\phi_{\textrm{peak}} = 21^{\circ}{_{-15}^{+17}}$, which agrees with the location and width of the flare excess observed in Figure \ref{fig:sdo_cdf_pdf}, suggesting that magnetic connectivity between active regions might favor sympathetic events. Note that we also tested a resolution of 360 seed points in longitude and 90 in latitude for the PFSS extrapolations across several Carrington rotations, but the results presented in Figure \ref{fig:pfss_distribution} remained unchanged.
\\\\
In conclusion, we have provided robust statistical evidence for the existence of sympathetic flares on the Sun based on observations from multiple instruments and using a significantly larger number of flares compared to previous studies. We find that sympathetic flares are unambiguously present on our Sun with an occurrence rate of about 5\%, and are generally found separated by an angle of about 30 degrees in longitude. However, further research is required to fully understand the underlying physical mechanisms of this phenomenon. An interesting avenue for this lies in the use of avalanche sandpile models, that have been used extensively in the past to study solar flare statistics (see \citealt{Charbonneau2001} for a comprehensive review) as well as for predictive purposes \citep[e.g.][]{StrugarekCharbonneau2014, Thibeault2022}. We are currently developing a numerical framework to test the efficiency of different types of inter-connectivity between active regions (modeled by different avalanching sandpiles), which results will be presented in a forthcoming paper.

\vspace{-0.1cm}
\section*{Data availability}
The catalog of identified sympathetic events for each instrument (SDO/AIA, \textit{RHESSI} and Solar Orbiter/STIX) is available online.\footnote{\url{https://github.com/lsguite/sympathetic_flares_catalogs.git}}
\\
\begin{acknowledgements}
L-SG acknowledges the financial support from the Natural Sciences and Engineering Research Council
of Canada (NSERC), the Fonds de recherche du Québec - Nature \& Technologies (FRQNT), a merit scholarship from Hydro-Québec, and an International Student Mobility Grant from the Center for Research in Astrophysics of Quebec (CRAQ). AS acknowledges support from the European Research Council (ERC) under the European Union’s Horizon 2020 research and innovation programme (grant agreement No 810218 WHOLESUN), from the Centre National d’Etudes Spatiales (CNES) Solar Orbiter, the Institut National des Sciences de l’Univers (INSU) via the Programme National Soleil-Terre (PNST), and the French Agence Nationale de la Recherche (ANR) project STORMGENESIS \#ANR-22-CE31-0013-01. PS acknowledges support from NSERC Discovery grant RGPIN-2024-04050. L-SG and PC are members of the CRAQ. We thank A. Finley for multiple discussions on flare databases and the heliospheric event knowledgebase. We thank L. Hayes for the production and maintenance of the curated STIX flare list.
\end{acknowledgements}

\bibliographystyle{aa}

\begin{appendix} \label{appendix}
\section{Solar flare catalogs}\label{sec:catalog}

\subsection{SDO/AIA}
We use the updated flare catalog from \citet{vanderSande2022}, which contains a total number of 16,270 flares observed by SDO/AIA that have a GOES X-ray magnitude C, M or X from 2010 to 2019. The AIA catalog is available on Github: \url{https://github.com/SWxTREC/aia-flare-catalog/tree/main/flare_catalogs}. Only flares triggered inside Space Weather HMI Active Region Patches (SHARPs; \citealt{Bobra2014}) with centers located within $\pm 65^{\circ}$ in heliographic coordinates from the solar center are included in the original list. We refer the reader to \citet{vanderSande2022} for a complete description of the construction of the flare list. These authors found that the SDO/AIA 131\AA \, filter correlated best with GOES X-ray fluxes, so we only include flares with spatiotemporal information at this wavelength in our study. This results in a final count of 13,263 flares.

\subsection{RHESSI}
The \textit{RHESSI} flare list contains 104,035 flares from 2002 to 2018 observed in 8 energy channels, available at: \url{https://hesperia.gsfc.nasa.gov/hessidata/dbase/hessi_flare_list.txt}. The positions of the flares are given in Helioprojective coordinates, so we perform a coordinate transformation to obtain the positions in both Heliographic Carrington and Heliographic Stonyhurst coordinates as well. We only retain flares triggered within $\pm 65^{\circ}$ in heliographic coordinates from the solar center that have an associated active region number. Due to \textit{RHESSI}'s orbit around the Earth, some flares are flagged for data gaps, eclipses, passage through the South Atlantic Anomaly (SAA), particle events, or count decimation during the flare interval. To ensure data quality, we only include flares that are not flagged, except for particle events that are not relevant to this study. This results in a final count of 20,172 flares. This filtering procedure is similar to that described in \citet{balazs2014}.

\subsection{Solar Orbiter/STIX}
We use the Solar Orbiter/STIX flare list, which is publicly available on GitHub: \url{https://github.com/hayesla/stix_flarelist_science}. The full procedure for generating this flare list, based on cleaned and filtered data from the operational STIX data center, is detailed on the same GitHub page. The list includes 17,031 flares, observed between 2021-03-18 and 2024-08-01, across all GOES classes; however, we filtered the list to include only flares with a GOES X-ray magnitude of C, M, or X. Additionally, we only include flares triggered within $\pm65^{\circ}$ of the limb, as observed from Solar Orbiter's vantage point at the time of the flare. This results in a final count of 8,306 flares. Note that not all flares were observed from Earth, depending on the position of Solar Orbiter.

\section{Distribution of active regions} \label{sec:AR}

\subsection{Active regions pairing}
\begin{figure}[h]
        \centering \includegraphics[width=\columnwidth]{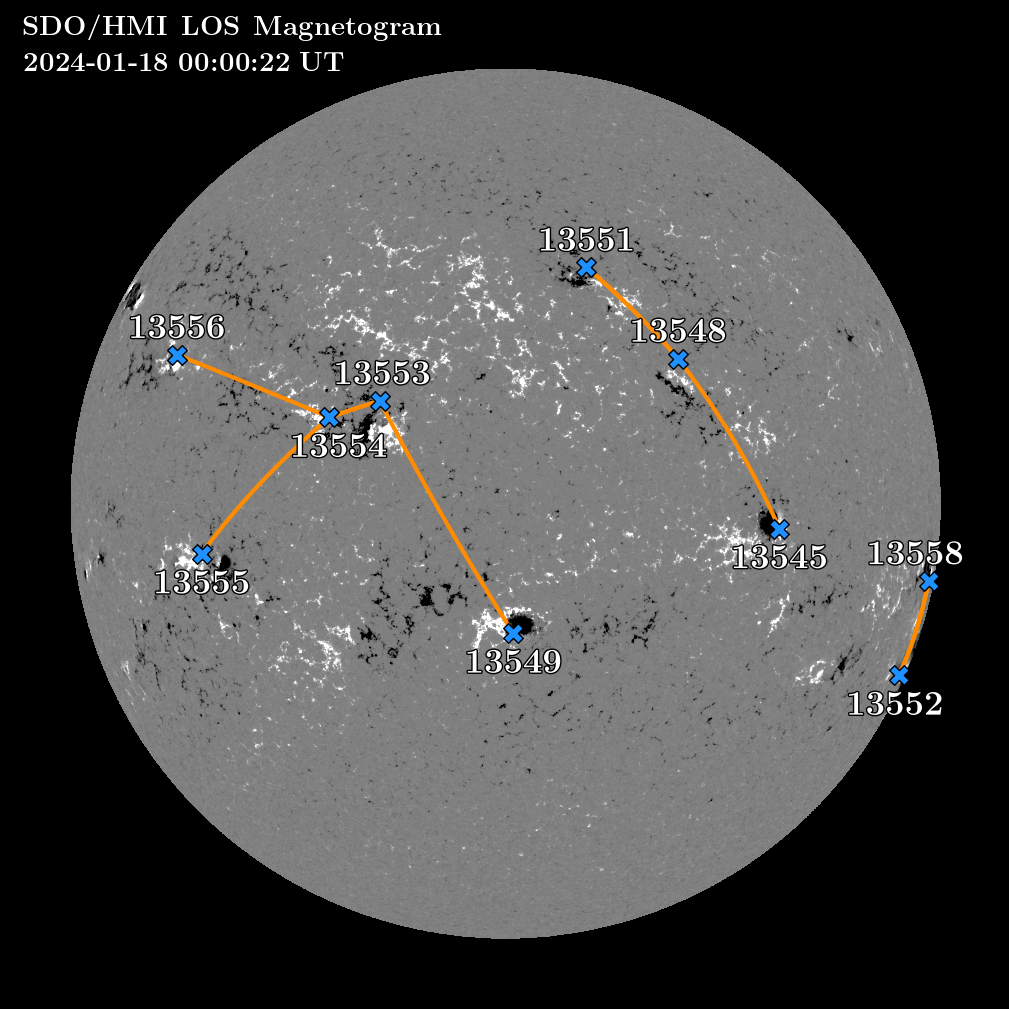}
        \caption{Identification of NOAA active regions on January 18, 2024, at 00:00:22 UT, marked by blue crosses on an SDO/HMI magnetogram. The orange lines represent the pairing of the closest active regions across both hemispheres, which is used to construct the distribution of their angular separation.}
        \label{fig:hmi}
    \end{figure}

To ensure that the overabundance of pairs of flares separated by $\Delta\phi \approx 30^\circ$ is not due to the longitudinal distribution of the active regions themselves, we construct the underlying spatial distribution of active regions on the solar disk for the period from 2002-01-07 to 2024-02-14. This period aligns with the time frames of the various flare lists used in this study (see Figure \ref{fig:SSN_TimeWindow}). For each Carrington rotation in this time interval, we query the Heliophysics Events Knowledgebase (HEK; \citealt{Hurlburt2012}) to retrieve the position, in heliographic Carrington coordinates, of all active regions that have a NOAA number associated with them. A single active region can have multiple data points over a Carrington rotation so we take the average spatial coordinate during this time interval. To validate this approximation, we calculate the standard deviation in longitude and latitude over all data points for each active region. For a total of 4937 active regions, the mean standard deviations are $\bar{\sigma}_{\phi} = 1.5^\circ$ in longitude and $\bar{\sigma}_{\theta} = 0.68^\circ$ in latitude. The small standard deviations justify using the average coordinates of each active region, given the required precision. Figure \ref{fig:hmi} provides a visual example of active region identification, with NOAA numbers labeled for each region marked by a blue cross on a magnetogram captured by the Helioseismic and Magnetic Imager (HMI; \citealt{Scherrer2012}) aboard SDO on January 18, 2024, at 00:00:22 UT.
    
\noindent For each NOAA active region (denoted as AR$_{i}$) observed during a given Carrington rotation, we identify the nearest active region (AR$_{k}$) that has an overlapping time window with AR$_{i}$. This ensures we compare active regions present on the solar disk simultaneously, rather than those that are spatially close but temporally different. An example of such pairing is shown in Figure \ref{fig:hmi}, where orange lines connect pairs of closest active regions marked by blue crosses (AR$_{i}$ and AR$_{k}$). We then measure the separation in latitude ($\Delta\theta$), longitude ($\Delta\phi$), and total angle ($\lambda$) between AR$_{i}$ and AR$_{k}$. This process is repeated across multiple Carrington rotations to increase the number of active regions in the sample. 

\begin{figure*}[h]
    \centering 
    \includegraphics[width = 2\columnwidth]{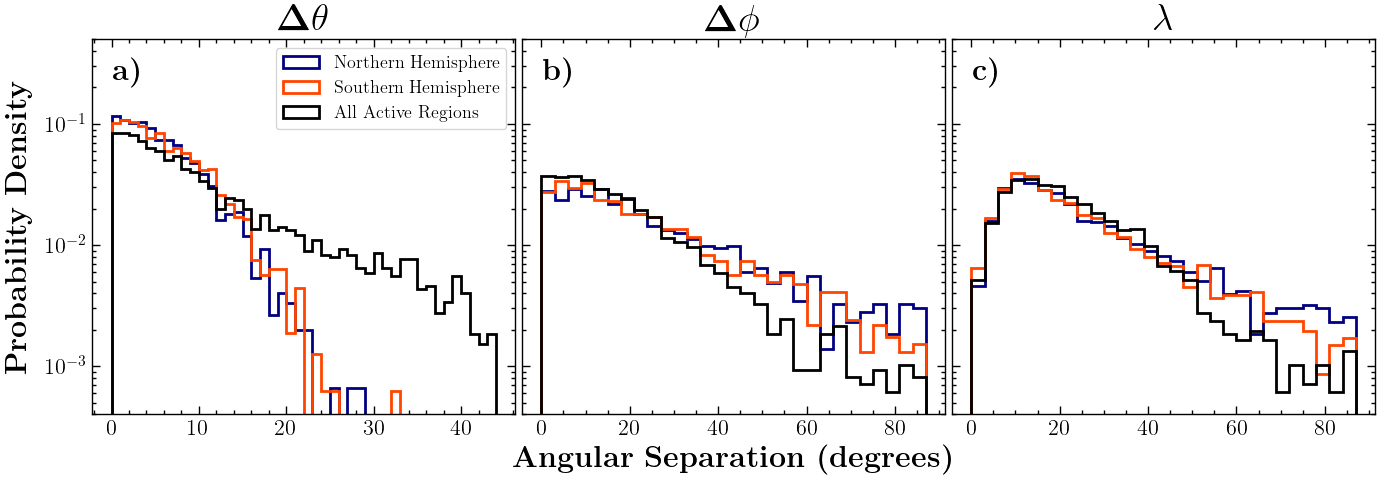}
    \caption{Distribution of the angular separation in latitude (a), longitude (b)  and total angle (c) between pairs of closest active regions overlapping in time on the solar disk. In each panel, we plot the distribution by considering active regions in the northern hemisphere only (blue), the southern hemisphere only (red) and all active regions from both hemispheres (black). The distributions are constructed using active regions listed in the Heliophysics Events Knowledgebase (HEK) that have an associated NOAA number for the period from 2002-01-07 to 2024-02-14 which covers the time frames of the various flare lists used in this study (see Figure \ref{fig:SSN_TimeWindow}).}
    \label{fig:hek_ar}
\end{figure*}

\begin{figure}[h]
    \centering 
    \includegraphics[width = \columnwidth]{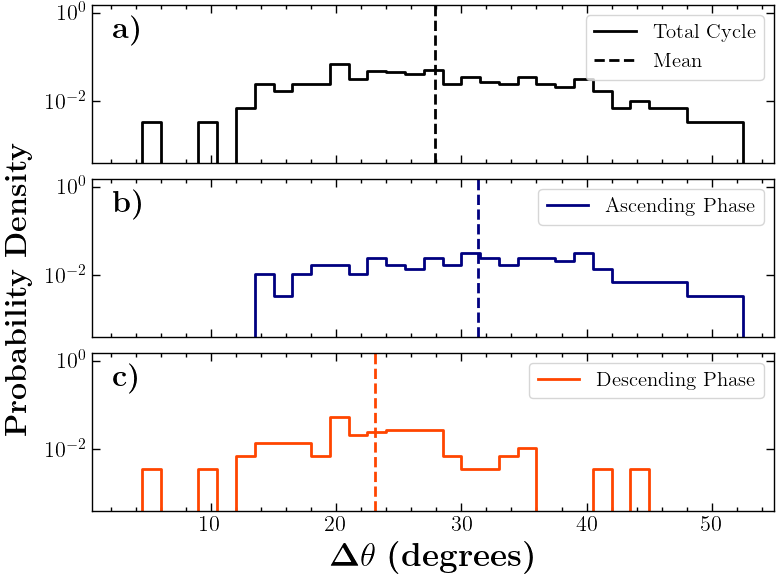}
    \caption{Distribution of angular separation in latitude between pairs of closest transequatorial active regions with $\Delta\phi \lesssim 5^{\circ}$. Panel (a) to (c) consider active regions present during nearly the entire solar cycle 24 (Carrington rotations 2098-2201), the rising phase (Carrington rotations 2098-2149) and the decaying phase (Carrington rotations 2150-2201), respectively. The time intervals approximately correspond to those indicated by the vertical black dashed lines in Figure \ref{fig:SSN_TimeWindow}. The vertical dashed lines indicate the distribution means.}
    \label{fig:hek_ar_transequatorial}
\end{figure}

\subsection{Global and hemispheric distribution of active regions}

\par The resulting distributions are shown in Figure \ref{fig:hek_ar}, with angular separation in latitude (panel a), longitude (panel b) and total angle (panel c). In each panel, we plot the distributions by considering active regions in the northern hemisphere only (blue), the southern hemisphere only (red) and all active regions from both hemispheres (black). Note that we consider every pair of nearest active regions to construct the angular separation distributions, as we expect flares to most significantly affect their nearby surroundings during sympathetic events. In panel (a), the combined-hemisphere histogram (black) shows an excess in the range $\Delta\theta \approx 20^{\circ} - 40^{\circ}$ compared to the hemispheric curves (red and blue), arising from pairings between active regions in the Sun's latitudinal activity bands across different hemispheres. As for the $\Delta\phi$ distribution in panel (b), all three curves follow an exponential trend and show no statistically significant excess of active regions separated by approximately 30$^{\circ}$, in contrast to the excess of flares observed in Figure \ref{fig:sdo_2d}. This suggests that the excess in the eruptive events likely comes from the flaring process itself and not the underlying distribution of active regions.

\subsection{Transequatorial distribution of active regions}
In Figure \ref{fig:transequatorial_mask}, we observe a deficit of consecutive transequatorial flares around $\Delta\theta \approx 25^{\circ}-30^{\circ}$ with $\Delta\phi \lesssim 5^{\circ}$. To determine if this deficit arises from the flaring process itself rather than the distribution of active regions, we conduct an analysis similar to that in Figure \ref{fig:hek_ar}, but by pairing the closest transequatorial active regions. Figure \ref{fig:hek_ar_transequatorial} shows the corresponding $\Delta\theta$ distribution, considering only active regions with $\Delta\phi \lesssim 5^{\circ}$. Panels (a) shows active regions covering Carrington rotations 2098-2201, representing nearly the entire solar cycle 24. Panels (c) and (d) display active regions during the rising phase (Carrington rotations 2098-2149) and the decaying phase (Carrington rotations 2150-2213) of cycle 24, respectively. The time intervals approximately correspond to those indicated by the vertical black dashed lines in Figure \ref{fig:SSN_TimeWindow}. Panels (b) and (c) show a gradual shift from larger to smaller $\Delta\theta$ separations, mirroring the trend in Figure \ref{fig:transequatorial_mask}(b-c), driven by the migration of active region emergence in the butterfly diagram. However, unlike Figure \ref{fig:transequatorial_mask}(a), we do not observe a deficit of active regions with $\Delta\theta \approx 30^{\circ}$ in Figure \ref{fig:hek_ar_transequatorial}(a). This suggests that the deficit of consecutive transequatorial flares in Figure \ref{fig:transequatorial_mask}(a) is caused by the flaring process and not the distribution of active regions.

\end{appendix}
\end{document}